\documentclass[aps,pra,twocolumn,superscriptaddress]{revtex4}%
\usepackage{graphics}
\usepackage[thmmarks]{ntheorem}
\usepackage{graphicx}
\usepackage{amsmath}
\usepackage{amsfonts}
\usepackage{amssymb}
\usepackage{float}
\usepackage{longtable}
\usepackage{epsfig}
\usepackage{latexsym}
\usepackage{theorem}
\usepackage{bbm}
\usepackage{bm}
\usepackage{psfrag}

\begin{document}
\title{Continuous-variable quantum key distribution in fast fading channels}
\author{Panagiotis Papanastasiou}
\affiliation{Computer Science and York Centre for Quantum Technologies, University of York,
York YO10 5GH, United Kingdom}
\author{Christian Weedbrook}
\affiliation{Xanadu, 372 Richmond St W, Toronto, M5V 2L7, Canada}
\author{Stefano Pirandola}
\affiliation{Computer Science and York Centre for Quantum Technologies, University of York,
York YO10 5GH, United Kingdom}

\begin{abstract}
We investigate the performance of several continuous-variable
quantum key distribution protocols in the presence of fading
channels. These are lossy channels whose transmissivity changes
according to a probability distribution. This is typical in
communication scenarios where remote parties are connected by
free-space links subject to atmospheric turbulence. In this work,
we assume the worst-case scenario where an eavesdropper has full
control of a fast fading process, so that she chooses the
instantaneous transmissivity of a channel, while the remote
parties can only detect the mean statistical process. In our study,
we consider coherent-state protocols run in various configurations,
including the one-way switching protocol in reverse reconciliation, the
measurement-device-independent protocol in the symmetric configuration
and a three-party measurement-device-independent network. We show
that, regardless of the advantage given to the eavesdropper (full
control of fading), these protocols can still achieve high rates.
\end{abstract}

\maketitle
\section{Introduction}
The purpose of quantum key distribution
(QKD)~\cite{Gisin2002,Scarani2008} is to establish a secret key
between two authenticated parties based on the laws of quantum
mechanics~\cite{NiCh}. This key can then be used for cryptographic
tasks such as the one-time pad~\cite{schneider}. In QKD, the
effect of eavesdropping on the exchanged quantum systems between
the parties can be detected and quantified so that a shared secret
key can be extracted. This is achieved by implementing the correct
amount of error correction and privacy amplification after
communicating via a public channel.
Significant advantages have been provided by the use of continuous
variable (CV) systems~\cite{RevModPhys.77.513}, in particular,
with Gaussian states~\cite{Weedbrook2012}. CV systems can transfer
higher amounts of information per signal with respect to
qubit-based approaches and they rely on cheaper technological
implementations. A number of CV-QKD protocols have been
studied~\cite{GG02,weedbrook2004noswitching,1way2modes,filip-th1,weed1,usenkoTH1,weed2,weed2way,usenkoREVIEW,pirs2way,2way2modes,1D-usenko}
and experimentally
implemented~\cite{1D-tobias,ulrik-Nat-Comm-2012,Grosshans2003b,jouguet2013,diamanti2007,ulrik-entropy,JosephEXP}.

In this scenario, another concept that needs to be treated
carefully is that of side channel attacks, where the eavesdropper
(Eve) creates an alternate channel with the aim of directly
attacking the setups where the signal states are prepared and
measured. A practical but partial solution was proposed in 2012
and known as measurement-device-independent (MDI)
QKD~\cite{side-channel attacks,MDILo}, later extended to the CV
setting~\cite{CV-MDI-QKD,CVMDIQKD-reply,Ottaviani2015}. An MDI-QKD
protocol can be seen as a ``prepare and measure'' version of
entanglement swapping~\cite{telereview} where the middle Bell
detection is performed by an untrusted relay. Recently, a suitable
generalization of the Bell detection to many parties has led
Ref.~\cite{starprotocol} to introduce a multipartite CV-MDI-QKD
star network with an arbitrary number of users.

Today the field of CV-QKD needs to accomplish two main
complementary tasks. The first one is the invention of practical
protocols that can achieve the high secret key rates that are
ideally accessible with CV systems. When implemented with ideal
reconciliation efficiencies, low-loss couplings and
highly-efficient detectors, CV-QKD protocols are not so far from
the ultimate Pirandola-Laurenza-Ottaviani-Banchi (PLOB) bound for
private communications over a lossy channel~\cite{PLOB15} (see
also extensions of this bound to multiple users~\cite{Multipoint},
repeaters and networks~\cite{networkPIRS}, and other
developments~\cite{nonPauli,FiniteSTRET,PirCosmo}). The other task
is improving the security analysis of CV-QKD protocols so as to
include realistic issues associated with their practical
realization, e.g., finite-size effects~\cite{rupertPRA,FS
CV-MDI-QKD}, and other aspects such as
composability~\cite{leverrierGEN,cosmoMDI-COMP}.

In terms of realistic implementations, one should also consider
the possibility of temporal variations of the communication line
between two remote users as modeled by the so-called fading
channel. In this case, the transmissivity $\eta$ of the link
between the two parties is not constant and may take values
according to some probability distribution~\cite{fading-channel}.
This description usually emerges from the fact that the parties
use a free-space link~\cite{laser in random media} that is
susceptible to the atmospheric turbulence~\cite{Atmospheric
channel 1,Atmospheric channel 2,Atmospheric channel 3,Atmospheric
channel 4,Atmospheric channel 5,Atmospheric channel 6,Atmospheric
channel 7}. Previous studies (e.g., see Ref.~\cite{CV-MDI-SAT})
have considered the symmetric situation where both users and
eavesdropper are subject to truly environmental fading. In this
manuscript, we consider a different situation, i.e., the
worst-case scenario where the eavesdropper is in complete control
of the quantum channel, so that she may choose different
instantaneous values of the transmissivity for each use of the
channel.

This type of fading is fast so that the users are only able to
estimate the statistical distribution of the transmissivity but
not its instantaneous values. This is in contrast to slow fading
where the transmissivity of the channel remains constant for
sufficiently many uses allowing the remote users to estimate its
actual value. In mathematical terms, for some fixed transmissivity
$\eta$ consider the key rate as given by the difference between
the mutual information $I_{\text{AB}}$ of the remote parties and
the accessible information $I_\text{E}$ of the eavesdropper, i.e.,
$R(\eta)=I_{AB}-I_E$. In slow fading, the key rate is averaged
over the distribution of the transmissivity. In fast fading, this
is not the most conservative approach. While we may still consider
the average $\tilde{I}_\text{E}$ for the eavesdropper, we need to
assume the lower transmissivity for the users, i.e.,
$I^{\eta_{\text{min}}}_{\text{AB}}$~\cite{note}. Therefore, the secret key
rate will be given by
\begin{equation}\label{Devetak-Winter}
R_{\text{fast}}=\beta
I^{\eta_{\text{min}}}_{\text{AB}}-\tilde{I}_\text{E},
\end{equation}
where $\beta\in [0,1]$ is a reconciliation parameter.

In this work, we adopt a basic model of fading channel where the
transmissivity is uniformly distributed over some interval. Then, we
study the performance of several CV-QKD protocols in the worst-case
scenario. We first investigate the effects in the case of the
one-way coherent-state switching protocol~\cite{GG02} in reverse
reconciliation (Sec.~\ref{one-way}). In Sec.~\ref{mdi}, we then
study the CV-MDI-QKD protocol~\cite{CV-MDI-QKD} in the symmetric
configuration~\cite{Ottaviani2015}. Finally, in Sec.~\ref{MP mdi},
we focus on the case of a CV-MDI-QKD network~\cite{starNETWORK}
considering three remote users. In all cases we show that high key
rates are achievable within reasonable distances, even in the
presence of fast-fading attacks.

\section{One-way QKD under fast fading \label{one-way}}
Consider a sender (Alice) preparing a bosonic mode $A$ using coherent
states whose mean values $\mathbf{x}=(x_q,x_p)$ are chosen
according to a zero-mean Gaussian distribution with variance
$\phi$.
\begin{figure}[t]
\vspace{+0.4cm}
\includegraphics[width=0.5\textwidth]{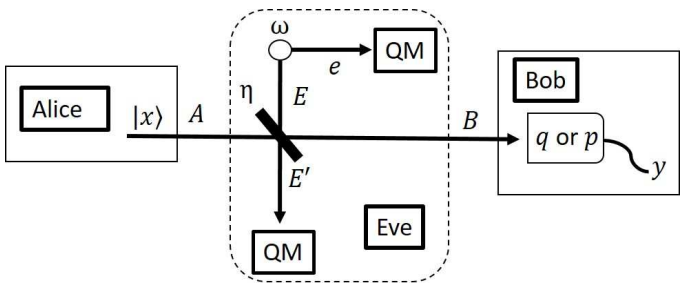}
\caption{\label{fig:one-way}One-way switching protocol. Alice
prepares a coherent state on mode $A$ whose mean value
$\mathbf{x}$ is modulated according to a Gaussian with variance
$\phi$. This state is sent through the channel with transmissivity
$\eta$ whose value may be changed by Eve in each use of the
channel. Bob gets an output mode $B$, which is homodyned randomly
in the $q$ or $p$ quadratures, with outcome $y$. Eve's attack also
comprises of her sending mode $E$ of an EPR state to interact with
mode $A$ in a beam splitter interaction with instantaneous
transmissivity $\eta$, and injecting thermal noise $\omega$. After
the interaction she stores mode $E'$ and the other EPR mode $e$ in
a quantum memory to be measured at the end of the quantum
communication (collective attack).}
\end{figure}
These states are sent through a channel with transmissivity $\eta$
to Bob, who then applies a homodyne measurement to either the $q$
or $p$ quadrature of his output mode $B$, with the outcome being
described by a random variable $y$ (see Fig.~\ref{fig:one-way}).
For each use of the channel, Eve has a two-mode squeezed vacuum
state \cite{Weedbrook2012} with thermal variance $\omega \geq 1$,
which represents a realistic version of an Einstein-Podolsky-Rosen
(EPR) pair in CV systems. This state describes modes $e$ and $E$
as depicted in Fig.~\ref{fig:one-way}. Then Eve's remote mode $E$
is made to interact with Alice's mode $A$ via a beam splitter with
transmissivity $\eta$ which takes values from a uniform
distribution $P_\eta$ with extremal values $\eta_{\text{min}}$ and
$\eta_{\text{max}}=\eta_{\text{min}}+\Delta\eta$. Thus, Eve
receives mode $E'$ and stores both $e$ and $E'$ in a quantum
memory, to be measured at the end of the entire quantum
communication.

An arbitrary input signal state $\rho$ undergoes a transformation
via a thermal-loss channel $\mathcal{E}_{\eta,\omega}(\rho)$ for a
specific $\eta$ randomly chosen by Eve (while $\omega$ is kept as
fixed). The asymptotic key rate will be given by
Eq.~(\ref{Devetak-Winter}). Here, Eve's information on Bob's
variable (reverse reconciliation) is given by the averaged Holevo
bound
\begin{equation}
\tilde{I}_\text{E}=\int d\eta P_\eta \chi(\text{E}:y),
\end{equation}
where
\begin{equation}
\chi(\text{E}:y)=S(\rho_{E'e})-S(\rho_{E'e|y})
\end{equation}
with $S(\cdot)$ being the von Neumann entropy computed over Eve's
output state $\rho_{E'e}$ and her conditional output state
$\rho_{E'e|y}$ (given Bob's outcome $y$).

The derivation is simplified by using the entanglement-based (EB)
representation of the protocol \cite{Weedbrook2012}, where, for
each use of the channel, Alice holds an EPR pair with parameter
$\mu=\phi+1$ and sends one of the modes through the channel. By
heterodyning her kept mode $a$, Alice projects the other travelling
mode $A$ into a modulated coherent state. This allows us to
exploit purification arguments and write
$S(\rho_{E'e})=S(\rho_{aB})$ and $S(\rho_{E'e|y})=S(\rho_{a|y})$.
\begin{figure}[t]
\includegraphics[width=0.52\textwidth]{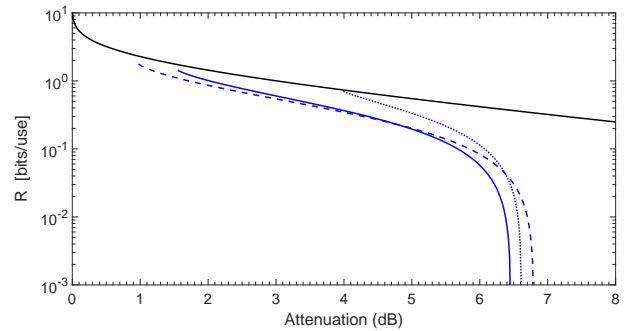}
\caption{\label{fig:plobbench}Fast fading channel. Secret key
rates are plotted for $\Delta\eta=0.2$ (dashed blue line),
$\Delta\eta=0.5$ (solid blue line) and  $\Delta\eta=0.6$ (dotted
blue line). We have set $\omega=1$ (passive eavesdropping),
$\beta=1$ (ideal reconciliation) and $\mu=10^6$. We compare the
results with the PLOB bound for repeaterless private
communications over a lossy channel (black line)~\cite{PLOB15}. We
can see that high rates can be achieved up to losses of about
$6-7$dB, where the rates start to rapidly decrease.}
\end{figure}
Furthermore, because the states involved are all Gaussian, we may
write the von Neumann entropy in terms of the symplectic
eigenvalues of the covariance matrices (CMs) of $\rho_{AB}$ and
$\rho_{A|y}$. In fact, for a Gaussian state whose CM has
symplectic spectrum $\{z\}$, we may write
\begin{equation}
S=\sum_z h(z),
\end{equation}
where
\begin{equation}
h(z)=\frac{z+1}{2}\log_2\frac{z+1}{2}-\frac{z-1}{2}\log_2\frac{z-1}{2}.
\end{equation}

\begin{figure}[t]
\vspace{-0.1cm} \centering
\includegraphics[width=0.52\textwidth]{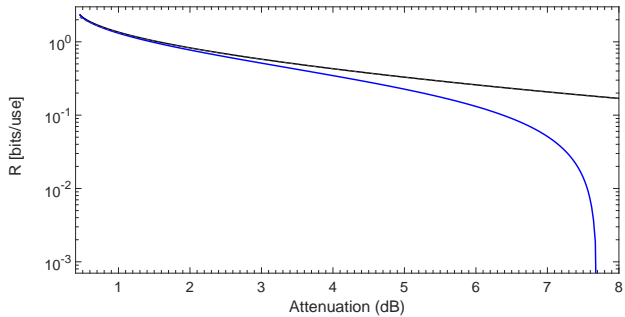}
\caption{\label{fig:owdB01}Comparison between fast and slow
fading. We plot the secret key rate for the fast fading channel
(lower blue line) and slow fading channel (upper black line) for
$\Delta \eta=0.1$. We also set $\mu=10^6$, $\omega=1$ (passive
eavesdropping) and $\beta=1$ (ideal reconciliation). Performances
are comparable within the range between $0$ and $6~{\rm dB}$.}
\end{figure}
\begin{figure}[t]
\centering
\includegraphics[width=0.52\textwidth]{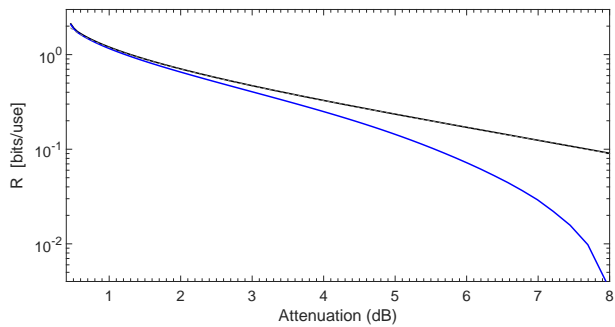}
\caption{\label{fig:owdB01w101b098}Comparison between fast and
slow fading. As in Fig.~\ref{fig:owdB01} but for $\omega=1.01$,
$\beta=0.98$, and optimized over $\mu$.}
\end{figure}

In Eq.~(\ref{Devetak-Winter}), the term
$I^{\eta_{\text{min}}}_{\text{AB}}$ represents the mutual
information between Alice and Bob, who does not know the
instantaneous value of the transmissivity but only the uniform
distribution $P_\eta$ with minimum transmissivity
$\eta_\text{min}$. Therefore, they need to choose the worst-case
scenario associated with the minimum possible transmissivity. As a
result, their mutual information turns out to be
$I_\text{AB}^{\eta_{\text{min}}}:=I_\text{AB}(\eta_{\text{min}})$
where
\begin{equation}
I_\text{AB}(\eta)=\frac{1}{2}\log_2 \frac{V_B}{V_{B|x_q}}.
\end{equation}
Here $V_B$ is the variance of Bob's variable $y$, and $V_{B|x_q}$
is the conditional variance given Alice's input $x_q$ (or $x_p$),
computed for a generic value of the transmissivity $\eta$.

In the regime of $\mu\gg1$ we may compute
\begin{equation}
\chi(\text{E}:y)=\frac{1}{2}\log_2\frac{ \eta
(1-\eta)\mu}{\omega}+h(\omega),
\end{equation}
and
\begin{equation}
I_\text{AB}(\eta)=\frac{1}{2}\log_2 \frac{
\eta\mu}{\eta+(1-\eta)\omega}.
\end{equation}
Thus, the secret key rate for a fast fading channel is given by
\begin{equation}
R_{\text{fast}}=\beta
I_\text{AB}(\eta_{\text{min}})-\frac{1}{\Delta\eta}\int^{\eta_\text{max}}_{\eta_\text{min}}
d\eta \chi(\text{E}:y).
\end{equation}
In particular, if we restrict ourselves to a pure-loss channel and
we set $\beta=1$ then the rate above simplifies to
\begin{align}\label{eq:ratefc}
R^\text{loss}_{\text{fast}}=&\frac{1}{2\Delta\eta}\left[g(\bar{\eta}_\text{min}-\Delta\eta)-g(\bar{\eta}_\text{min})\right]\nonumber\\
-&\frac{1}{2\Delta\eta}(\eta_\text{min}+\Delta\eta)\log_2\frac{\eta_\text{min}+\Delta\eta}{\eta_\text{min}}\nonumber\\+&\log_2e,
\end{align}
where $g(x)=x\log_2x$ and $\bar{x}=1-x$.
Note that, for slow fading, both Alice and Bob's mutual
information and Eve's Holevo information need to be averaged so
that
\begin{equation}
R_{\text{slow}}=\frac{1}{\Delta\eta}\int^{\eta_\text{max}}_{\eta_\text{min}}
d\eta [\beta I_\text{AB}(\eta)-\chi(\text{E}:y)].
\end{equation}



In Fig.~\ref{fig:plobbench} we show the secret key rate for a
fast-fading channel with $\Delta\eta=0.2$, $\Delta\eta=0.5$ and
$\Delta\eta=0.6$, also compared with the PLOB bound, which sets
the limit for repeaterless private communication over a lossy
channel~\cite{PLOB15}. In Fig.~\ref{fig:owdB01} we compare the key
rates for slow and fast fading considering $\Delta \eta=0.1$. In
Fig.~\ref{fig:owdB01w101b098}, we consider the secret key rates
for $\Delta \eta=0.1$ but including extra thermal noise
$\omega=1.01$ and assuming a non-ideal reconciliation parameter
$\beta=0.98$ (rates are optimized over $\mu$). As we can see from
the plots, the key rate is high up to losses of the order of
$6-7$dB, even in the presence of fast fading attacks.
\section{CV-MDI-QKD under fast fading \label{mdi}}
The detailed calculations for the CV-MDI-QKD protocol can be found
in the Supplementary Material of
Ref.~\cite{CV-MDI-QKD,Ottaviani2015}. Here we consider the
symmetric configuration, so that each link with the untrusted
relay is a fading channel whose transmissivity ($\eta_A$ and
$\eta_B$) follows a uniform probability distribution, while the
thermal noise $\omega$ is equal and fixed. For fast fading, we
have
\begin{align}
R_{\text{fast}}^{\text{MDI}}=&\beta I_{\text{AB}}(\eta_{\text{min}})-\nonumber\\
&-\frac{1}{(\Delta
\eta)^2}\int^{\eta_{\text{min}}+\Delta\eta}_{\eta_{\text{min}}}
\int^{\eta_{\text{min}}+\Delta\eta}_{\eta_{\text{min}}} d \eta_A d
\eta_B \chi(\eta_A,\eta_B),
\end{align}
where $I_{\text{AB}}(\eta)$ and $\chi(\eta_A,\eta_B)$ are given in
Refs.~\cite{CV-MDI-QKD,Ottaviani2015}.

\begin{figure}[t]
\vspace{-0.1cm} \centering
\includegraphics[width=0.52\textwidth]{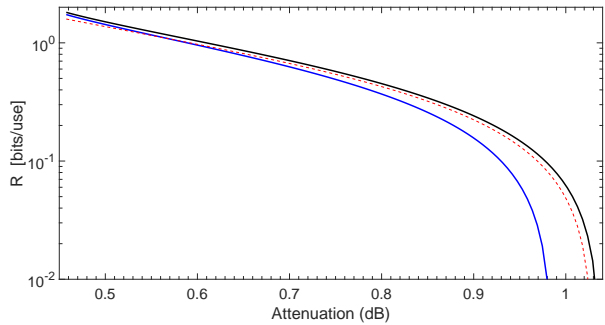}
\caption{\label{fig:mdidb01}Performances of the CV-MDI-QKD
protocol in symmetric configuration assuming two fading channels
in the links with $\Delta \eta=0.1$ and no excess noise
($\omega=1$). We plot the secret key rate for fast fading (lower
blue line), slow fading (upper black line), and also for the
standard case of a non-fading lossy channel (middle red dashed
line) over the expectation value of $\eta$, i.e., $\bar{\eta}=
\eta_{\min}+\frac{\Delta\eta}{2}$. We set $\beta=1$ and
$\mu=10^6$.}
\end{figure}

\begin{figure}[t]
\includegraphics[width=0.52\textwidth]{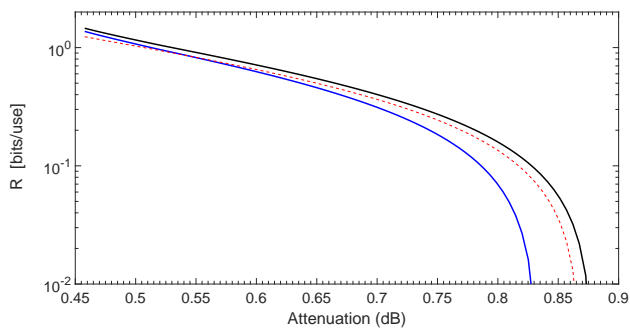}
\caption{\label{fig:mdidb01w101correlatedb098}We plot the same
cases as in Fig.~\ref{fig:mdidb01}, but assuming a two-mode
correlated attack with noise $\omega=1.01$, imperfect
reconciliation $\beta=0.98$ and optimization over $\mu$. As
expected, the key rates deteriorate.}
\end{figure}

In Fig.~\ref{fig:mdidb01}, we present secret key rates for
$\beta=1$, $\omega=1$ and very large modulation $\mu\simeq10^6$,
while we set $\Delta\eta=0.1$. We see that the performance for
fast fading is not so far from that related to slow fading and
that is achievable with a standard lossy channel. In
Fig.~\ref{fig:mdidb01w101correlatedb098}, we then present the same
instances but for $\beta=0.98$, optimizing over $\mu$ and setting
$\omega=1.01$. In this latter case, the eavesdropper may also
optimize her attack by exploiting correlations in the injected
environmental state~\cite{CV-MDI-QKD}.

\section{CV-MDI-QKD three-user network under fast fading\label{MP mdi}}
\begin{figure}[t]
\vspace{0.4cm}
\includegraphics[width=0.5\textwidth]{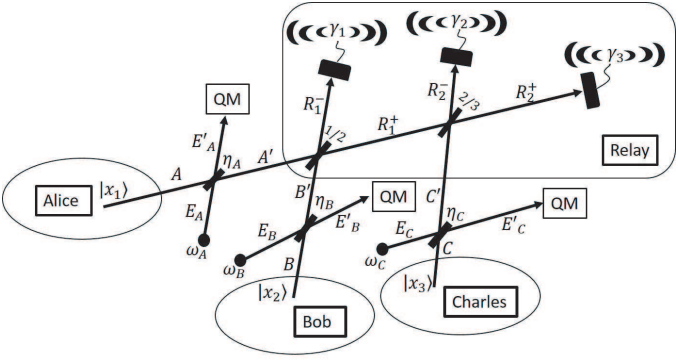}
\caption{\label{fig:tri-party} Three-party CV-MDI-QKD network. The
parties prepare coherent states in the modes $A$, $B$ and $C$
whose mean values are Gaussian variables $\mathbf{x}_1$,
$\mathbf{x}_2$ and $\mathbf{x}_3$, with variance $\phi$. Then the
parties send these states to the relay using links with
transmissivities $\eta_A$, $\eta_B$ and $\eta_C$ respectively.
After traveling through the links the modes arrive at the relay as
modes $A'$, $B'$ and $C'$ and processed by the relay. Although the
relay is under the full control of the eavesdropper, we can assume
without loss of generality that it operates consistently in each use
of the channel: (a) firstly it mixes Alice's and Bob's modes in a
beam splitter with transmissivity $1/2$ and measures the
$q$-quadrature of mode $R^-_{1}$ with a homodyne detection, (b)
subsequently mixes Charlie's mode with $R^+_{1}$ and then measures
the $q$-quadrature and $p$-quadrature  of modes $R^-_{2}$ and
$R^+_{2}$ respectively, (c) finally the results of the
measurements $\gamma_1 $, $\gamma_2$ and $\gamma_3$ are broadcast.
As in Ref.~\cite{CV-MDI-QKD}, any general attack affecting both
the links and the relay can be reduced to an attack tampering only
with the links. In this case, Eve is injecting thermal noise
$\omega_1$, $\omega_2$ and $\omega_3$ in each of the links by
means of the modes $E_A$, $E_B$ and $E_C$ interacting with modes
$A$, $B$, $C$. In a general Gaussian attack, Eve's modes are
described by a correlated Gaussian state whose covariance matrix
is specified in Eq.~\eqref{eq:threepartyattack}.}
\end{figure}

Let us start with investigating the protocol assuming lossy
channels. We consider a three-party network where Alice, Bob and
Charlie prepare coherent states for their modes $A$, $B$ and $C$.
The mean values are Gaussian variables $x_1$, $x_2$, and $x_3$
with the same variance $\phi$. Then they send these states to an
untrusted relay through three links described by lossy channels
with transmissivities $\eta_A$, $\eta_B$ and $\eta_C$
respectively. The relay is assumed to operate in a certain way in
each channel use. In particular, as illustrated in
Fig.~\ref{fig:tri-party}, it mixes Alice's and Bob's modes in a
beam splitter with transmissivity $\tau_1=1/2$ and homodynes the
$q$ quadrature of the output mode $R^-_1$, while the mode $R^+_1$
is mixed with Charlie's mode in a beam splitter with
transmissivity $\tau_2=2/3$. Then, the output modes $R^-_2$ and
$R^+_2$ are homodyned with respect to the $q$ and $p$ quadrature,
respectively. All the measurement results
$\boldsymbol{\gamma}=(\gamma_1,\gamma_2,\gamma_3)$ are then
broadcast (see Fig.~\ref{fig:tri-party}). This is the three-party
realization of the multipartite CV Bell detection recently
introduced in Ref.~\cite{starNETWORK}. The distributed classical
correlations can be processed by the three parties to derive a
common secret key for secure quantum conferencing. See
Ref.~\cite{starNETWORK} for a CV-MDI-QKD quantum conferencing
network with arbitrary number of parties (and Ref.~\cite{Ribeiro}
for a recent fully device-independent quantum conferencing network
in discrete variables).

Although the relay is assumed to be under the control of the
eavesdropper, we can always assume an attack that is described by
an attack restricted in the links according to the discussion in
Ref.~\cite{CV-MDI-QKD}. More specifically, a correlated attack
among all the three links is described by a  covariance matrix.
Here each of the modes are interacting by a beam splitter with
each of the modes that the parties send through the channel. This
CM is given by
\begin{equation}\label{eq:threepartyattack}
\mathbf{V}_{E_A E_B E_C}=\begin{pmatrix}
\omega_A \mathbf{I}&\mathbf{G}_1&\mathbf{G}_3\\
\mathbf{G}_1&\omega_B \mathbf{I}&\mathbf{G}_2\\
\mathbf{G}_3&\mathbf{G}_2&\omega_C \mathbf{I}
\end{pmatrix},
\end{equation}
where $\omega_A$, $\omega_B$ and $\omega_C$ are the vectors of the
noise injected by the eavesdropper in each link whereas
$\mathbf{G}_i=\text{diag}(g_i,g_i^\prime)$ describes the
correlations between the modes. When $g_i$ and $g^\prime_i$ are
equal to zero, the attack is reduced to an uncorrelated attack,
which is the case that we investigate in this study.

In terms of the security analysis, we adopt the EB representation
of the protocol, where the (traveling) modes  $A$, $B$ and $C$ are
each one half of an EPR pair with parameter $\mu=\phi+1$. Then
heterodyne measurements are applied to the ancillary EPR modes
$a$, $b$, and $c$ so that the traveling modes are projected onto
modulated coherent states. In this representation, Eve's Holevo
bound $\chi$ is given by the symplectic eigenvalues of the total
CM $\mathbf{V}_{abc|\boldsymbol{\gamma}}$ and the conditional CM
$\mathbf{V}_{bc|\boldsymbol{\gamma},x_1}$ following the reasoning
in Sec.~\ref{one-way}. In particular, the total CM is defined as
the CM of the parties' local modes after the application of the
three relay measurements with outcomes $\boldsymbol{\gamma}$, and
the conditional CM is derived by the total CM after applying a
heterodyne detection on mode $a$ (we assume that Alice's variable
is the one to reconciliate with).

\begin{figure}[t]
\includegraphics[width=0.5\textwidth]{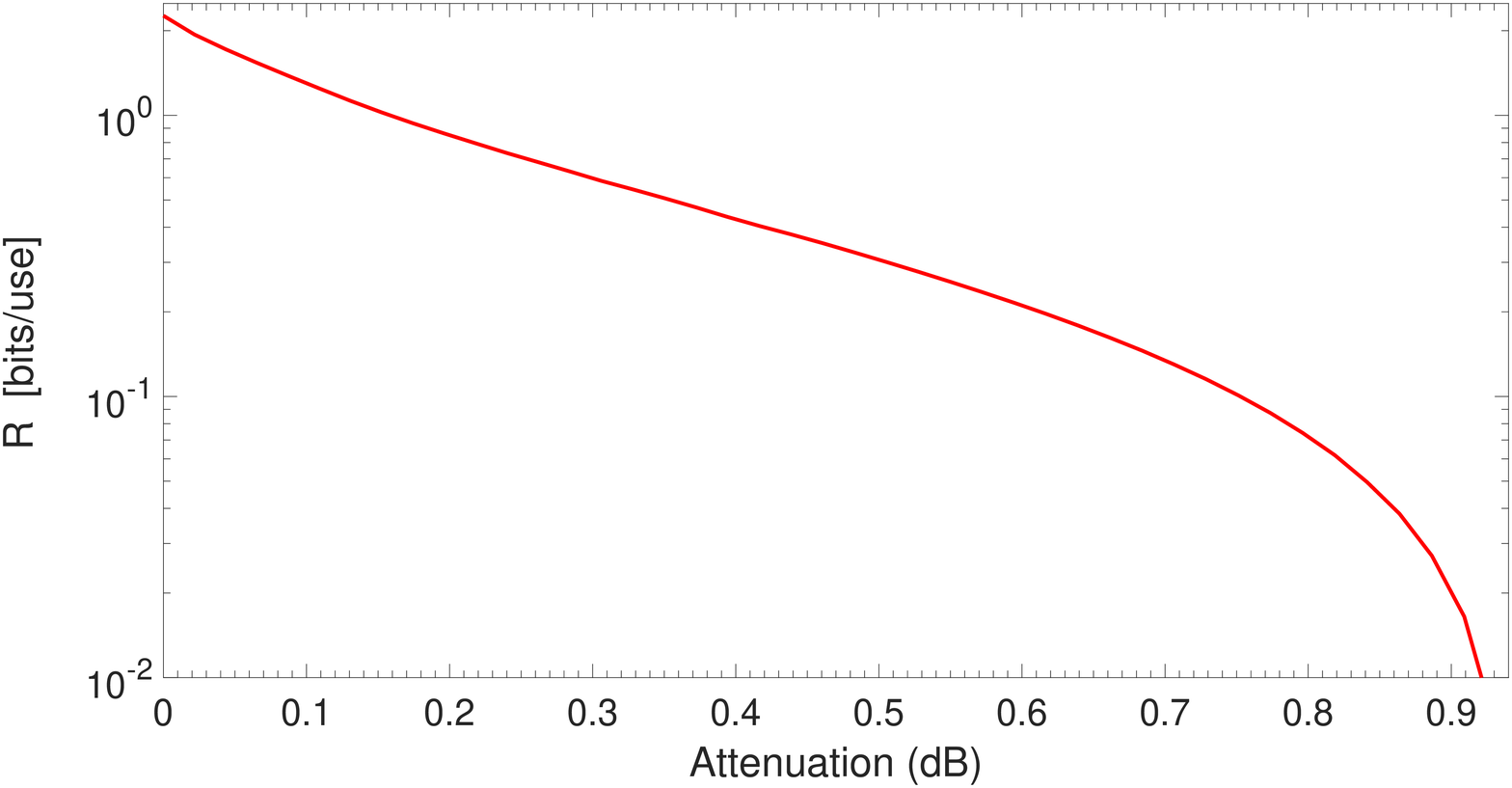}
\caption{\label{fig:dB3partystandard} The rate for the three-party
star configuration protocol based on the CV-MDI-QKD scheme,
optimized over $\mu$.}
\end{figure}

The mutual information in the case of the three parties is defined
as the minimum of the mutual information between Alice-Bob and
Alice-Charlie, i.e., $I_{\text{min}}=\min\{I_{AB},I_{AC}\}$. Each
of the terms are evaluated by the formula
$I_{AB(C)}=\frac{1}{2}\log_2\boldsymbol{\Sigma}_{b(c)}$, where we
have the following (for $m=b$ or $c$)
\begin{equation}
\boldsymbol{\Sigma}_m=\frac{\text{det}(
\mathbf{V}_{m|\boldsymbol{\gamma}})+\text{tr}(
\mathbf{V}_{m|\boldsymbol{\gamma}})+1}{\text{det}(
\mathbf{V}_{m|\boldsymbol{\gamma},x_1})+\text{tr}(
\mathbf{V}_{m|\boldsymbol{\gamma},x_1})+1},
\end{equation}
in terms of the covariance matrices of the local mode $m$.

As a result, the secret conferencing key rate is given by
\begin{equation}\label{eq:3mdirate}
R(\mu,\boldsymbol{\omega},\boldsymbol{\eta})=\beta
I_{\text{min}}(\mu,\boldsymbol{\omega},\boldsymbol{\eta})-\chi(\mu,\boldsymbol{\omega},\boldsymbol{\eta}),
\end{equation}
where $\boldsymbol{\omega}=(\omega_A,\omega_B,\omega_C)$ is the
vector of the noise injected by the eavesdropper in each link with
corresponding transmissivity
$\boldsymbol{\eta}=(\eta_A,\eta_B,\eta_C)$. Numerically, we have
checked that, even for ideal reconciliation $\beta=1$, the largest
value of $\mu$ is not the optimal and we have therefore to
optimize the rate over $\mu$. In the case of a star configuration, the
previous rate simplifies to
\begin{equation}\label{eq:3mdiratestar}
R^\text{star}(\mu,\omega,\eta)=\beta
I_{\text{min}}(\mu,\omega,\eta)-\chi(\mu,\omega,\eta),
\end{equation}
where $\eta_A=\eta_B=\eta_C:=\eta$ and
$\omega_A=\omega_B=\omega_C:=\omega$. This is plotted in
Fig.~\ref{fig:dB3partystandard} for passive eavesdropping
($\omega=1$).

\begin{figure}[t]
\includegraphics[width=0.5\textwidth]{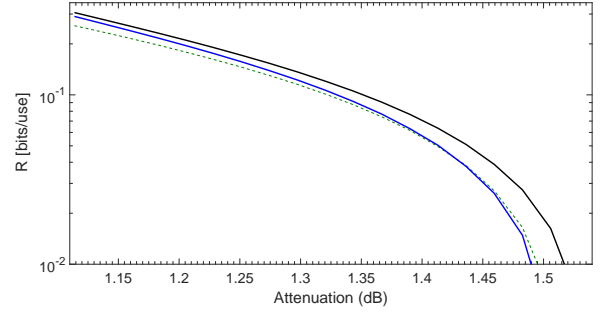}
\caption{\label{fig:dB3partydh005} The secret conferencing key
rate in a star configuration of the three-party CV-MDI-QKD network
assuming fast-fading channels (blue solid line) and slow-fading
channels (black solid line) with the same variance $\Delta
\eta=0.05$. We also include the rate of the protocol in the
presence of lossy channels with transmissivity
$\bar{\eta}=\eta+\frac{\Delta\eta}{2}$. For all the plots we have
optimized over $\mu \in [2,20]$ and set $\omega=1$.}
\end{figure}

Consider the star configuration in the presence of fading channels
affecting the links, with uniform distribution between
$\eta_{\text{min}}$ and $\eta_{\text{min}}+\Delta \eta$. We need
to integrate the Holevo bound with respect to the three
transmissivities of the channels and compute the mutual
information assuming the minimum transmissivity (worst-case
scenario). Thus, we write
\begin{equation}
R_{\text{fast}}^{\text{star}}=\beta
I_{\text{min}}(\mu,\omega,\eta_{\text{min}})-\iiint^{\eta_{\text{min}}+\Delta\eta}_{\eta_{\text{min}}}
\frac{\chi(\mu,\omega,\boldsymbol{\eta})}{(\Delta\eta)^3}
d\boldsymbol{\eta},
\end{equation}
The rate for $\Delta\eta=0.05$ and $\omega=1$ is optimized over
$\mu$ and shown in Fig.~\ref{fig:dB3partydh005}. From the figure,
we can see that the performance is comparable to the case of
slow-fading where the parties' mutual information is averaged over
the statistical distribution.

\section{Conclusion}
In this work we have investigated the effects of fading channels in
the links used by authorized parties in various quantum key
distribution protocols. More specifically, we have studied the one-way
switching protocol with coherent states in reverse
reconciliation, the symmetric configuration of continuous-variable measurement-device-independent quantum key distribution
protocol, and its extension to a three-user network for quantum
conferencing. Fading describes channels with randomly varying
transmissivity according to a probability distribution that
encompasses effects present in free-space communications where the
links are under the influence of atmospheric turbulence. Here,
we have considered the most random scenario where the
distribution is uniform between two extremal values.

In particular, our work considers the worst-case scenario where
the eavesdropper is assumed to have the full control of the fading
channel. In other words, it is Eve who fixes the instantaneous value
of the transmissivity, not just the environment. When this value
is changed very fast, e.g., for each transmission, then we have a
fast fading attack which makes the honest user in a particularly
disadvantage situation. They can only access the probability
distribution of fading at the end of the quantum communication and
they therefore need to assume the minimum transmissivity
(compatible with that distribution) for the extraction of their
secret key.

As we discussed in our paper, this is clearly different from a
slow fading attack where the action of the eavesdropper is slow
with respect to the quantum communication so that the
transmissivity is approximately constant over a large block size.
This allows the honest user to make an estimate of the
transmissivity to be used in the extraction of part of the key. In
any case, our work shows that the performance achievable in the
worst-case scenario in the presence of fast fading is not so far
from the performance under slow fading. In particular,
sufficiently high rates can be achieved within ranges of distance
which are typical of the various protocols analyzed. Such results
prove the robustness of continuous-variable quantum key distribution protocols under conditions of
turbulence which may be typical in realistic free-space scenarios.

\section{Acknowledgements}
C.W. would like to acknowledge the Office of Naval Research
program Communications and Networking with Quantum
Operationally-Secure Technology for Maritime Deployment
(CONQUEST), awarded to Raytheon BBN Technologies under prime
contract number N00014-16-C-2069.  The content of this paper does
not necessarily reflect the position or policy of the Government
and no official endorsement should be inferred. P. P. acknowledges
support from the EPSRC via the `UK Quantum Communications Hub'
(EP/M013472/1).
\appendix


\begin{thebibliography}{99}
\bibitem{note} The approach is similar to what happens
in a finite-size security analysis~\cite{rupertPRA,FS CV-MDI-QKD},
where the step of parameter estimation provides a statistical
distribution for the values of the channel parameters. From this
statistical distribution, the parties choose worst-case values for
the parameters, e.g., a minimum value of the transmissivity that
is compatible with the observed distribution. In the same fashion,
in the presence of a fast fading channel, there is a probability
distribution associated with the transmissivity and the most
conservative strategy for Alice and Bob is to choose the
worst-case value. In particular, for a uniform distribution, this
means to pick the minimum value in the interval of possible
transmissivities.
\end{thebibliography}
\end{document}